\documentclass[apj]{emulateapj}
\submitted{Accepted by ApJ, February 18, 2005}

\newcommand{\Lose}{\ensuremath{L_{\mathrm{178}}}}
\newcommand{\Ledd}{\ensuremath{L_{\mathrm{Edd}}}}
\newcommand{\Medd}{\ensuremath{\dot M_{\mathrm{Edd}}}}
\newcommand{\Mbh}{\ensuremath{M_{\mathrm{BH}}}}
\newcommand{\Msun}{\ensuremath{M_{\sun}}}
\newcommand{\mdot}{\ensuremath{\dot m}}
\newcommand{\Mdot}{\ensuremath{\dot M}}
\newcommand{\mcrit}{\ensuremath{\dot m_{\mathrm{crit}}}}
\newcommand{\Lbol}{\ensuremath{L_{\mathrm{bol}}}}
\newcommand{\eps}{\ensuremath{\epsilon}}
\newcommand{\epslo}{\ensuremath{\eps^{\mathrm{lo}}}}
\newcommand{\epshi}{\ensuremath{\eps^{\mathrm{hi}}}}
\newcommand{\LMplane}{$(\Mbh,\Lbol)$ plane}

\shorttitle{A simple test for two accretion modes in AGN}

\begin{document}
\title{A simple test for the existence of two accretion modes\\ in
Active Galactic Nuclei}

\author{Sebastian~Jester}
\affil{Fermilab, MS 127, PO Box 500, Batavia IL, 60510, USA}
\email{jester@fnal.gov}

\begin{abstract}
By analogy to the different accretion states observed in black-hole
X-ray binaries (BHXBs), it appears plausible that accretion disks in
active galactic nuclei (AGN) undergo a state transition between a
radiatively efficient and inefficient accretion flow.  If the
radiative efficiency changes at some critical accretion rate, there
will be a change in the distribution of black hole masses and
bolometric luminosities at the corresponding transition luminosity.
To test this prediction, I consider the joint distribution of AGN
black hole masses and bolometric luminosities for a sample taken from
the literature.  The small number of objects with low Eddington-scaled
accretion rates $\mdot < 0.01$ and black hole masses $\Mbh <
10^9\,\Msun$ constitutes tentative evidence for the existence of such
a transition in AGN.  Selection effects, in particular those
associated with flux-limited samples, systematically exclude objects
in particular regions of the \LMplane. Therefore, they require
particular attention in the analysis of distributions of black hole
mass, bolometric luminosity, and derived quantities like the accretion
rate.  I suggest further observational tests of the BHXB-AGN
unification scheme which are based on the jet domination of the energy
output of BHXBs in the \emph{hard} state, and on the possible
equivalence of BHXB in the \emph{very high} (or \emph{steep
power-law}) state showing ejections and efficiently accreting quasars
and radio galaxies with powerful radio jets.
\end{abstract}


\section{Introduction}\label{s:intro}

It has recently been shown that there is a common scaling relation
between X-ray luminosity, radio luminosity and black hole mass in
black-hole X-ray binaries (BHXBs) and active galactic nuclei (AGN),
strongly suggesting that the accretion mechanism in both classes of
objects is identical \citep{MHD03,FKM04}.  Given this universality of
accretion physics, it is reasonable to assume that the different
observational states of BHXB should also exist in AGN
\citep*{Mei01,MGF03}. In this paper, I suggest a simple test looking
for observational consequences of such an equivalence of accretion
states.

The observational states of BHXBs are distinguished by different
spectral shapes and, to some extent, different luminosities.  They are
identified with different physical modes of accretion onto a compact
object \citep[see the review by][and references therein; these authors
  suggest a non-standard state nomenclature which will be used
  here]{MR04}.  One marked transition between accretion states occurs
at Eddington ratios of order 1-10\%, between the ``\emph{hard X-ray,
  steady jet}'' (conventionally called ``\emph{low (luminosity)/hard
  (spectrum)}'') state, and the ``\emph{thermal-dominant}''
(``\emph{high/soft}'') state.  The difference between the \emph{hard}
and \emph{thermal-dominant} observational states is ascribed to a
difference in the state of the underlying accretion disk around a
stellar-mass black hole.  The \emph{thermal dominant} state is
identified with radiatively efficient accretion through a standard
\citet{SS73} disk, while the \emph{hard} state is identified with a
radiatively inefficient accretion flow that replaces the innermost
part of the standard disk at low accretion rates \citep*[RIAF;
  e.g.,][and references therein]{Esiea97}.  A second transition occurs
at Eddington ratios of 20--30\%, where the standard \citet{SS73} thin
disk becomes unstable.  There, the spectrum changes from being
dominated by thermal emission to showing a ``\emph{steep power-law}''
(\emph{very high state}), i.e., an X-ray spectrum that is harder than
a pure blackbody but softer than that of the \emph{hard} state.  This
transition may also be linked to a decrease in the radiative
efficiency and hence a smaller thermal contribution to the total
emission.  Alternatively, an increased contribution from inverse
Compton scattering in a magnetically heated corona may be responsible
for the harder spectrum \citep{HaMa91}.

The link between the state of the accretion disk and presence or
absence of a jet is an important part of the picture currently
emerging for BHXBs and its analogy in AGN.  As implied by its name,
the \emph{hard/steady-jet} state is the only accretion state for which
continuous, steady jets have been observed (and quite possibly are
always present).  When an object enters the \emph{thermal-dominant}
state, the radio emission associated with the steady jet is quenched
\citep*[][e.g.]{GFP03}. Similarly, there are suggestions that the
radio galaxies and quasars with powerful jets are analogues of
black-hole binaries in the \emph{steep power-law} state \citep*[or
making the transition to that state; see][]{FBG04} which show
transient X-ray and radio flares, interpreted as discrete ejections of
high-velocity jets \citep{Mei01,GFP03}.

While the detailed physics of accretion flows are subject of ongoing
research and debate, the generic feature of the models for the
transition between the \emph{hard} and \emph{thermal dominant} state is a
change in the radiative efficiency at some critical value of $\mdot$,
the mass accretion rate \Mdot\ normalized to the Eddington accretion
rate.  This transition occurs at low accretion rates $\mdot = \mcrit \approx
0.02$.  I show here that it is a consequence of such a state
transition that there is a change in the distribution of accretion
flow luminosities at the critical accretion rate $\mdot=\mcrit$
(\S\ref{s:theory}).  Using data from the literature, I examine whether
the distribution of observed AGN black hole masses and luminosities is
consistent with this aspect of a state transition in accretion flows
(\S\ref{s:obs}).  The theoretical and observational picture of the
second state transition, from \emph{thermal dominant} to \emph{steep power
law}, is much less clear.  Therefore, I concentrate on the
low-efficiency transition in the first part of the paper. However, the
existence of this accretion state may be the key to resolving some
apparent contradictions of the simplest unification picture, which are
discussed in \S\ref{s:disc}.  I conclude in \S\ref{s:conc}.

\section{Effect of a radiative efficiency change on the joint distribution of
luminosities and black hole masses}\label{s:theory}

In this section, I show that a transition from a radiatively
inefficient accretion state (a RIAF in general) to one with much
higher radiative efficiency (a radiatively efficient accretion flow
[REAF] such as a standard \citealt{SS73} disk) is expected to change
the density of objects in the $(\Mbh,\Lbol)$ plane at the luminosity
corresponding to the transition accretion rate \mcrit.  I assume the
simplest possible generic model of an accretion state transition
occurring at low radiative efficiency.  The notation used here follows
\citet{Esiea97}, but the argument is germane to any transition from
RIAF to REAF, whatever the detailed physical theory of either state
may be.  Thus, even if the \emph{form} of the radiative efficiency is
in actuality different from that used here, the \emph{method} of
identifying the difference between different accretion states by
considering the distribution of objects in the $(\Mbh,\Lbol)$ plane is
universally applicable.

The accretion flow radiates a fraction $\zeta$ of the rest mass of
accreted matter, so that the radiative bolometric luminosity \Lbol\
and the physical accretion rate \Mdot\ are related by
\begin{equation}
\Lbol = \zeta \Mdot c^2
\label{eq:lbol_phys}
\end{equation}
In accretion disk models, the radiative efficiency does not depend
directly on \Mdot, but only on the dimensionless accretion rate $\mdot
= \dot M / \Medd$ \citep[see, e.g.,][]{Cheea95,Esiea97}.  In this
case, using the accretion-rate dependent efficiency in
Equation~\ref{eq:lbol_phys} makes the definition of \Medd\ circular,
because \Medd\ depends on the efficiency, which depends on \mdot,
which depends on \Medd.  Therefore, if the radiative efficiency varies
with accretion rate, it is necessary to define \Medd\ for a
\emph{fixed} fiducial value of the radiative efficiency. This is
usually chosen as $\zeta_\mathrm{fid}=0.1$ (\citealt{NY95,Esiea97},
but compare \citealt{Cheea95,BB99,MCF04}, e.g., who use
$\zeta_\mathrm{fid}=1$; $\mdot=1$ corresponds to $\Lbol=\Ledd$ if
\emph{and only if} the same value is used for the efficiency in the
calculation of $\Mdot$ and $\Medd$ from $\Lbol$ and $\Ledd$,
respectively).  \Medd\ is then given by
\begin{eqnarray}
\Medd & = & \Ledd/(0.1 c^2) \nonumber\\
      & = & 2.2\times10^{-8} \frac{\Mbh}{\Msun}\; \Msun
      \mathrm{yr}^{-1}
\label{eq:theory.medd}
\end{eqnarray}
For all purposes other than calculating \Medd, I separate the
conversion efficiency $\zeta$ into $\zeta = \eps\,\eta$.  Here, $\eta$
is the dissipation efficiency, the fraction of the rest mass energy of
accreted matter that is liberated by the accretion process.  This
energy may either be radiated away, swallowed by the black hole as
thermal or kinetic energy, or converted into kinetic energy of a disk
wind or jets \citep{NY95,BB99,FGJ03}.  The fraction that is radiated away is denoted by
\eps. Since accreting matter has to lose the same binding energy in
both presumed accretion states, I assume that universally
$\eta=0.1$. The power liberated by the accretion flow is thus a
constant $P = 0.1 \Mdot c^2$, while the radiated luminosity is given
by
\begin{eqnarray}
\Lbol & = & \eps \mdot \, \Ledd \nonumber \\
      & = & \eps \mdot \frac{\Mbh}{\Msun} \, 1.26\times 10^{31}\,\mathrm{W}. \label{eq:defeps}
\end{eqnarray}
In the accretion state transition scenario, \eps\ is a property of the
accretion flow solution and therefore depends on \mdot: for $\mdot
\leq \mcrit$, \eps\ is the low radiative efficiency of a RIAF, while
at $\mdot > \mcrit$, it is the high efficiency of a REAF.
\citet{Esiea97} found $\mcrit = 0.08$ based on the ``strong ADAF''
proposal \citep{NY95}, while \citet*{MLM00} predict $\mcrit = 0.02$
based on their coronal evaporation model.  Observations of black-hole
binaries find a value close to $\mcrit = 0.02$ \citep{Mac03}, For pure
numerical convenience, we use $\mcrit = 0.01$ here; any conclusions
that would be affected by the difference between $\mcrit = 0.01$ and
$\mcrit = 0.02$ would be on a weak footing anyway. Again ignoring
physical details, I assume the following simple form for the radiative
efficiency as function of accretion rate:
\begin{equation}
  \eps = \left\{ \begin{array}{ll}
     \epslo = 100 \mdot & \mathrm{ if\ } \mdot \leq \mdot_{\mathrm{crit}} \\
     \epshi = 1  & \mathrm{ if\ } \mdot_{\mathrm{crit}} < \mdot.
	 \end{array}
    \right.
\label{eq:valeps}
\end{equation}
The form of $\epslo$ follows \citet{NY95}.  The numerical factor 100
in the definition of \epslo\ is chosen to make the luminosity
continuous as \mdot\ crosses \mcrit, since transitions between the
spectral states can occur without a large change in luminosity.  As
mentioned in the introduction, the accretion rate \mdot\ should in
principle be restricted to $\mdot < 0.3$ since the standard
\citet{SS73} thin disk becomes unstable and BHXBs are known to enter
the \emph{steep power-law} state at higher accretion rates.  Lacking
accurate knowledge of the radiative efficiency of that state, it
appears more appropriate to concentrate on the detectability of the
low-accretion rate transition, which should not be influenced by the
presence of this third accretion state at high accretion rates (see
\S\ref{s:disc.superEdd}).

The assumed prescription for the radiative efficiency in
Equation~\ref{eq:valeps} ignores the finding that there is a
hysteresis in the transition luminosity in at least five soft X-ray
transient binary systems, in the sense that the transition from the
\emph{hard} to the \emph{thermal dominant} state occurs at a
luminosity that is higher than a factor of about five than the reverse
transition (see \citealp{MC03} and references therein;
\citealp{BO02}).  I will discuss the possible impact of this
simplification in \S\ref{s:disc.crit.parval}.

It is immediately obvious from equations \ref{eq:defeps} and
\ref{eq:valeps} that for fixed black hole mass, the distribution of
bolometric luminosities must show the same transition to a different
scaling with \mdot\ at $\mdot=\mcrit$ as \eps. Thus, this transition
should be detectable when considering the distribution of accreting
systems in the \LMplane, as long as the distribution of accretion
rates \mdot\ does not similarly change precisely at \mcrit. This
assumption is consistent with the common view that \mdot\ is an
object-specific input parameter fixed by the surroundings of the
particular accreting system (e.g., the availability of gas in the AGN
host galaxy), while \mcrit\ is a universal property of the accretion
flow solution \citep[although it is possible that the mass supply rate
  at large radii does not in fact govern the accretion rate onto the
  black hole; see][e.g.]{Pro05}.  Thus, I consider whether there any
observational evidence for such a change in the distribution of
observed AGN luminosities at fixed black hole mass.

\section{Observations}\label{s:obs}

\subsection{Results}

\begin{figure}
\plotone{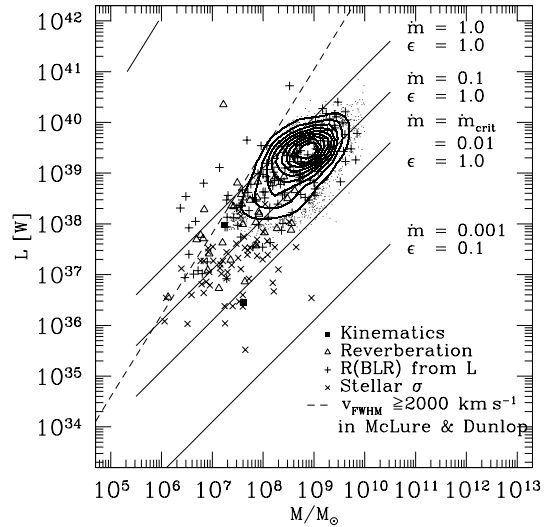}
\epsscale{1}
\caption{Distribution of AGN bolometric luminosities and black hole
  masses (the symbol type indicates the method of black hole mass
  determination as detailed below).  The diagonal solid lines are
  lines of constant Eddington ratio $\Lbol/\Ledd$. They show the
  bolometric luminosity as function of black hole mass from Equations
  \protect\ref{eq:defeps} and \protect\ref{eq:valeps} for the given
  values of the accretion rate \mdot. The radiative efficiency \eps\
  has a discontinuity at $\mdot=\mcrit=0.01$ which separates
  radiatively inefficient flows (RIAF, $\eps = 100 \mdot$) at smaller
  \mdot\ from radiatively efficient flows (REAF, $\eps = 1$) at larger
  \mdot.  The diagonal lines are equally spaced in \mdot, so that the
  change in line spacing with respect to luminosity should be
  reflected in a change in the density of objects in the \LMplane\ at
  the line $\mdot=\mcrit$.  Contours and dots are measurements for the
  12245 objects from the SDSS DR1 quasar catalog \citep{SFHea03},
  determined by \citet{MD04} using the relation between optical
  luminosity and size of the broad-line region (BLR) together with the
  line width of the H\,$\beta$ or \ion{Mg}{2} lines as virial mass
  indicators.  These authors compute bolometric luminosities using a
  fixed bolometric correction for the optical luminosity.  There are
  12245 objects in total; the contours indicate the density of objects
  in bins of 0.15\,dex in both axes, starting at 1 object per bin and
  increasing by a factor of $\sqrt{2}$ per contour. The dashed line
  indicates the FWHM cut $\geq 2000$\,km\,s$^{-1}$ applied by
  \citeauthor{MD04} which excludes objects above and to the left of
  this line. The solid line in the upper left-hand corner shows the
  correlated change in \Lbol\ and \Mbh\ that an error in an object's
  optical luminosity of 1\,dex would produce.  Other symbols are those
  black hole mass measurements from \citet{WU02a} for which the
  authors have determined the bolometric luminosity. The method of
  black hole mass determination is indicated by the symbol type: solid
  squares for resolved stellar kinematics (2 objects); open triangles
  for reverberation mapping (36); plus signs for BLR size from optical
  luminosity as in \citeauthor{MD04} (139); and crosses for direct
  measurements of the stellar velocity dispersion $\sigma$ and the
  \Mbh-$\sigma$ relation (57).
\label{f:obs}}
\end{figure}
Figure~\ref{f:obs} shows the distribution of bolometric luminosities
and black hole masses for the sample of objects from the compilation in
\citet{WU02a}, who list black hole masses obtained by a variety of
methods, and from \citet{MD04}, who determine virial black hole masses
for objects from the Sloan Digital Sky Survey (SDSS) quasar catalog
\citep{SFHea03} using the virial method and the correlation between
AGN luminosity and size of the broad-line region (BLR).  The SDSS
spectroscopic quasar survey is an optically flux-limited survey of
quasar candidates selected either from 5-color photometry or as
optical point-source counterparts of sources detected in the FIRST
radio survey \citep[details of the selection are described
in][]{qso_ts}.  The objects whose mass and luminosity are taken from
\citet{WU02a} are taken from a variety of samples and sources.  Most
of the sources with black hole masses from the BLR size-luminosity
relation are radio-selected quasars, while most other AGN are Seyferts
in nearby galaxies, or optically selected quasars from the Bright
Quasar Survey (PG objects).

The diagonal lines in Fig.\,~\ref{f:obs} indicate the relation between
luminosity and black hole mass for the indicated Eddington-scaled accretion
rates $\mdot$ (equally spaced in $\log \mdot$) and the
resulting radiative efficiency from Equation~\ref{eq:valeps}.  As
argued above, the distribution of objects in $\mdot$ is not expected
to change at $\mdot=\mcrit$. Therefore, the change in radiative
efficiency at $\mdot=\mcrit$ should lead to a change in the density of
objects in the \LMplane\ at the line $\mdot=\mcrit$.  In
Fig.\,\ref{f:obs}, the density of objects does indeed change abruptly
at this line.  In fact, nearly all objects from the samples used here
lie \emph{above} this line in the radiatively efficient regime, while
only a few objects have lower $\mdot$ as inferred from their black
hole mass, luminosity, and the radiative efficiency prescribed by
Equation~\ref{eq:valeps}.

\begin{figure}
\plotone{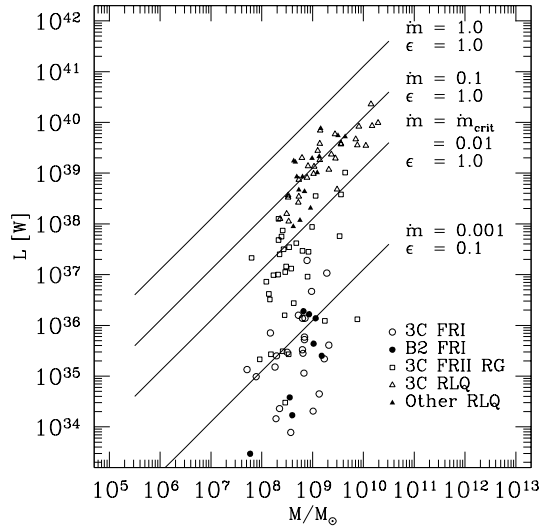}
\caption{Bolometric luminosity against black hole mass for sources
  listed by \citet{MCF04}, including only sources with a measurement
  of the bolometric luminosity (excluding upper limits).  The lines
  for different Eddington-scaled accretion rates are identical to
  those in Figure~\ref{f:obs}.  Triangles indicate radio-loud quasars
  (empty triangles for 3CR objects, filled for objects from
  \citealp{MD01} and \citealp{HFHea02}), squares are FR~II radio
  galaxies from the 3C sample, and circles are low-luminosity (mostly
  FR~I) radio galaxies from the 3CR (empty) and B2 (filled) samples.}
\label{f:3cobs}
\end{figure}
Figure\,\ref{f:3cobs} shows the equivalent of Fig.\,\ref{f:obs} for
the sample of radio-loud objects from \citet*{MCF04}.  That sample
includes only objects with detected host galaxies and HST imaging,
allowing the authors to measure the black hole mass from the
correlation between host galaxy luminosity and black hole mass, as
well as the nuclear optical luminosity to determine the nuclear
bolometric luminosity using a bolometric correction factor.  I have
omitted sources with only an upper limit to the nuclear
luminosity. Some of the objects classified as radio-loud quasars are
also included in the sample shown in Figure~\ref{f:obs}. The accretion
rate histogram derived from this figure (essentially, by a projection
of this figure on a line perpendicular to the lines
$\mdot=$\emph{const}), \citet{MCF04} shows a clear bimodality in the
accretion rate distribution of AGN.  This bimodality is taken as
strong evidence for the presence of a transition between two accretion
states.  The presence of a ``gap'' in the distribution, roughly
located at $3\times10^{-4} < \mdot < 10^{-2}$, requires a
discontinuous change of some property of the accretion flow (note that
in contrast to the treatment here, \citealt{MCF04} define $\mdot
\equiv \Lbol/\Ledd$, i.e., they use a constant radiative efficiency to
avoid having to make assumptions about the radiative efficiency as
function of accretion rate).  Even though the accretion rate histogram
clearly shows a bimodality, there is no obvious structure to the
distribution of the same objects in the \LMplane.  In fact, it appears
preferable to consider the full 2-dimensional distribution rather than
its 1-dimensional projection, the accretion rate histogram, because
the impact of selection effects is more obvious in the 2-dimensional
diagram.  The impact of selection effects is discussed in detail in
\S\ref{s:obs.selection} below.

The bulk of the 12245 objects from \citet{MD04} lies in the region
between $\mdot=0.1$ and $\mdot=1$ (contours in
Fig.\,\ref{f:obs}). These are so far from the line $\mdot=\mcrit$ that
these objects genuinely cannot be accreting in the
low-\mdot\ radiatively inefficient mode.  This suggestion seems to be
supported by the fact that the vast majority of SDSS quasars has a
blue bump \citep[the average SDSS quasar spectrum clearly shows the
  big blue bump; see][]{Berea01,YCVBea04}.  However, the implied
\mdot\ of many of these objects is so high that the \citet{SS73}
standard disk solution does not hold any more and they should be in
the third accretion state.  We return to this point in the discussion
section below (\S\ref{s:disc.spec.higheff}).  The lack of
inefficiently accreting objects in the SDSS sample is predominantly
due to the combined effect of the luminosity and flux limit for
objects in the SDSS quasar catalog, from which this sample is drawn
(see \S\ref{s:obs.selection} below).  The lack of such objects
therefore does not allow to make any statement about the presence or
absence of two accretion modes in AGN.

The objects from \citet{WU02a} include many Seyfert galaxies with
luminosities that are much lower than the limit for inclusion in the
SDSS quasar catalog, in particular among the objects with \Mbh\ from
stellar velocity dispersions (crosses in Fig.\,\ref{f:obs}) and those
with reverberation-mapping masses (triangles).  These objects are
preferentially found above the line $\mdot=\mcrit=0.01$.  The lack of
objects with $\mdot < \mcrit$ at $\Mbh > 10^{9}\Msun$ again is mainly
due to selection effects. But even at $\Mbh < 10^{9}\Msun$, there are
fewer objects with $\mdot < \mcrit$ compared to $\mdot > \mcrit$ (12
below \mcrit\ and 185 above).  The asymmetry is perhaps most
significant among the objects with masses from host galaxy stellar
velocity dispersions, which are more likely to include low-luminosity
AGN: only 11 of these are below \mcrit, compared to 46 above.  The
significance of this difference is hard to quantify because the
objects listed by \citet{WU02a} by construction do not constitute a
complete sample.  But the small number of objects with $\mdot <
\mcrit$ at $\Mbh < 10^{9}\Msun$ may be taken at least as
weak suggestive evidence for the presence of a radiatively inefficient
accretion mode.

In stark contrast to the small number of low-efficiency objects in
Fig.~\ref{f:obs}, the \citet{MCF04} sample shown in
Fig.\,\ref{f:3cobs} shows a substantial number of objects in the
low-efficiency regime.  The key difference is that \citet{MCF04} used
only optical observations and a bolometric correction to determine
bolometric luminosities for these low-power radio galaxies, while
\citet{WU02a} explicitly excluded radio galaxies in their
determinations of bolometric luminosities because ``obscuration and
beaming are significant'' in these sources.  By contrast,
\citet{MCF04} argue \citep[based on the analysis by][]{CMSea02} that
nuclear obscuration is not important in these low-power radio
galaxies, so that the nuclear optical luminosity can be used to infer
the bolometric luminosity.  This finding is disputed by \citet{CR04}
--- while they agree with \citet{CMSea02} that the optical core flux
of these objects is dominated by jet emission, they conclude that the
jet emission comes from scales larger than the obscuring material.
Thus, I refrain from drawing any conclusion from the large number of
low-luminosity (and hence low-accretion rate) sources in
Figure\,\ref{f:3cobs}, since they would hinge on which assumptions are
adopted about the nature of the nuclei of low-luminosity radio
galaxies.

I next consider the impact of the other assumptions going into
construction of the distributions in the \LMplane\ on the results
obtained here.

\subsection{Impact of assumptions}
\label{s:obs.assumptions}

In order to assess the significance of any conclusions about the
presence or absence of two accretion modes, it is necessary to
consider the impact of the assumptions going into the determination of
the distribution of objects in the \LMplane.  Perhaps the most serious
caveat has already been mentioned above: the accretion rate
\mdot\ might not necessarily be the same as the mass supply rate, but
could be set by processes internal to the accretion disk
\citep{Pro05}, so that the value of \mdot\ may be a property of the
accretion disk solution rather than an independent parameter set by
the availability of gas supply in the host galaxy, as has been assumed
here. If this is the case, the distribution of objects in the
\LMplane\ would still provide information about the physics of
accretion disks if the mass supply rate could be determined
independently.

A detailed discussion in \S\ref{s:obs.err} of the appendix shows that
errors in the determination of black hole mass and bolometric
luminosities would either blur the transition region or systematically
shear the distribution of points in the \LMplane, making the detection
of the transition more difficult, but not impossible.  In
\S\ref{s:obs.lum} of the appendix, I conclude that the bolometric
luminosity constitutes the most reliable measure of accretion
luminosity.  However, large systematic errors are introduced in
bolometric luminosities by deriving bolometric corrections from one
sample \citep[the objects in the library of AGN SEDs by][]{EWMea94}
but applying them to the SDSS quasar sample, which goes beyond
traditional UV excess selection and includes objects with much redder
optical colors, and therefore possibly different SEDs in other
wavelength regions, than the samples on which the bolometric
corrections are based.

In the remainder of this section, I consider in detail the impact of
assumptions about the values for critical accretion rate and radiative
efficiency, and of the sample selection function on this distribution.
I will then discuss the implications of the observed distributions.

\subsection{Values for critical accretion rate and radiative
efficiencies}\label{s:disc.crit.parval}

In order to determine the accretion rate \mdot\ from the observable
\Lbol, it was necessary to assume a value for the fraction of the
accreted matter's rest-mass energy that is dissipated by the accretion
process, and to assume radiative efficiencies for the different
accretion states.  In addition, the location of the line dividing
efficient from inefficient accretors in the \LMplane\ obviously
depends on the value of \mcrit.

Here, the fraction of rest-mass energy that is liberated as accretion
power was taken to be $\eta=0.1$ universally.  However, this value in
fact depends on the location of the last stable orbit around the black
hole, which in turn depends the black hole spin.  For a non-rotating
black hole (Schwarzschild metric), $\eta=0.06$.  The efficiency of
accretion on a rotating black hole (Kerr metric) can be lower than the
Schwarzschild value if it is counter-rotating with respect to the
disk, while a maximally co-rotating black hole would result in
$\eta=0.42$.  As there is likely a large variation of the black hole
spin from AGN to AGN based on its particular accretion history, the
value of $\eta$ will also differ from object to object.  The value of
$\eta$ changes the location of the lines $\mdot = \mathrm{const.}$ in
Fig.\,\ref{f:obs}, i.e., there would be a different set of these lines
for each value of $\eta$.  The appropriate simplification would be to
use the ensemble average of $\eta$ to draw this set of lines; since an
accurate measurement of $\eta$ is not available for any object, use of
the conventional value $\eta=0.1$ appears most appropriate.

I have chosen \mcrit, \epslo, and \epshi\ to match observations of
black-hole binaries approximately.  If the true values of \mcrit, \epslo,
and \epshi\ were smaller than assumed here, the transition between
efficient and inefficient accretion would occur at smaller
luminosities, where there are not many objects in Fig.\,\ref{f:obs}.
In this case, \emph{all} objects in Fig.~\ref{f:obs} might in fact be
in the efficient state, and the luminosity range occupied by
inefficiently accreting objects has not yet been reached.

Another complication arises from the fact that there is be a
hysteresis loop between the two accretion states in black-hole
binaries, with the low-high transition occurring at a different
accretion rate than the high-low transition \citep[][and references
therein]{MC03}. These authors find that the transition back to the
low-efficiency hard state occurs at a luminosity $\sim5$ times lower
than the hard-to-soft transition from low to high efficiency, so that
objects with identical luminosity can have different accretion rates.
If AGN can undergo an accretion state change during their active
phase, this would apply to them, too, so that the instantaneous
luminosity would not reveal the instantaneous accretion rate.  The
adopted value of \mcrit\ would again need to be interpreted as an
ensemble average.

In all these cases, the transition between efficient and inefficient
accretion would still be visible in the \LMplane, but the transition
would be blurred by the scatter of true efficiencies about the assumed
average value, or the location of the transition would be moved. In
this case, it might occur in a region that currently does not contain
any objects.  These difficulties can therefore be overcome with
observations of more objects.

\subsection{Ignoring a possible third accretion state}\label{s:disc.superEdd}

Even if the parameters chosen for the low-accretion rate transition
are correct, there is a third \emph{super-Eddington} accretion mode
with a different efficiency of the standard thin disk. In fact, the
standard thin-disk model becomes self-inconsistent and must be
modified to include advection above $\mdot=$0.2--0.3 \citep[e.g.]{KB99},
i.e., they become radiatively inefficient, too.  Accretion at even
higher rates could become even more radiatively inefficient
\citep[see][e.g.]{BM82,Abr04}.  ``Super-Eddington'' would then refer
to the fact that the accretion rate exceeds the Eddington \emph{rate}
for the fiducial efficiency, while the radiative \emph{luminosity} may
still be \emph{below} the Eddington luminosity because of the decrease
in the actual radiative efficiency.  Consideration of this accretion
state is particularly relevant since BHXBs enter a \emph{steep
power-law} state at roughly 30\% of the Eddington luminosity, with a
harder X-ray spectrum than the thermal-dominated state.

The physics of these high-accretion rate inefficient disks are the
subject of ongoing research.  No universally accepted predictions are
available yet for their radiative efficiency and possible
time-dependent behavior.  Generically, though, the radiative
efficiency must be dropping with increasing \mdot\ just above this
second critical accretion rate. The accretion luminosity would
therefore increase more slowly than \mdot. It could even
\emph{decrease} with increasing \mdot\ if the accretion efficiency
scales like $\mdot^p$ with $p<-1$ in the transition region.  The same
radiative luminosity may then be produced either by a standard
accretion flow with modest \mdot\ and high \eps, or by a
super-Eddington flow with high \mdot\ and modest \eps.  In any case, a
second transition to a low-efficiency state would lead to a ``pileup''
of sources in the \LMplane\ near the super-Eddington \mcrit. In this
context, it may be of relevance that the contours for the SDSS quasars
in Fig.~\ref{f:obs} are peaked around the line $\mdot=0.3$.

In an extreme case, the radiative efficiency of the super-Eddington
state could be so low that the luminosity of objects with \emph{any}
\Mbh\ can be produced in a radiatively inefficient flow. In this case,
there may in fact be \emph{no} standard radiatively efficient
accretion flows at all \citep[cf.][]{Cheea95}.  Unless this is the
case, this high-Eddington rate transition at $\mdot=0.2$--0.3 should
not influence the detectability of the low-Eddington rate transition at
$\mcrit\approx 0.01$.

\subsection{Impact of sample selection effects}
\label{s:obs.selection}

The selection of a sample of AGN for which black hole masses and
absolute luminosities are to be determined always restricts the
resulting distribution of objects to certain regions of the
\LMplane. This can happen both explicitly, by a luminosity or line
width cut, e.g., or implicitly, because the distribution of AGN
luminosities and black hole masses over redshift systematically
removes objects from flux-limited surveys, e.g.  The correct
interpretation of the observed distribution of objects in the
\LMplane\ requires to take into account these selection effects.

\subsubsection{Selection effects in \citet{MCF04} sample}
\label{s:obs.sel.mcf}

\begin{figure}
\plotone{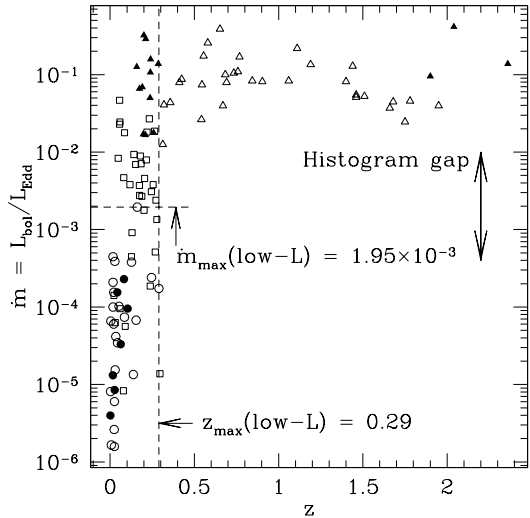}
\epsscale{1}
\caption{\label{f:MCFmdotz}Dimensionless accretion rate \mdot\ against
 redshift $z$ for the \citet{MCF04} objects shown in
 Figure\,\ref{f:3cobs}; for consistency with that work, this figure
 defines $\mdot \equiv \Lbol/\Ledd$, i.e., with a constant radiative
 efficiency.  Symbols as in Figure\,\ref{f:3cobs}: triangles for radio-loud
 quasars, empty triangles for 3CR and filled from two other samples;
 squares for 3CR FR~II radio galaxies; circles for low-luminosity
 radio galaxies from 3CR (empty) and B2 (filled).  The double arrow
 indicates the location of the gap identified by \citet{MCF04} in the
 accretion rate histogram.  The dashed lines show the maximum value of
 \mdot\ and $z$ for the set of low-luminosity radio galaxies (open
 circles).  These constitute the majority of low-\mdot\ objects, but
 are drawn from a completely different volume than the 3CR radio-loud
 quasars (open triangles), which constitute the majority of
 high-\mdot\ objects: the former are limited to $z\leq 0.29$ and
 $\mdot \leq 1.95\times10^{-3}$, while the latter are found at
 $z\geq0.3$ and $\mdot > 10^{-2}$.  Figure\,\ref{f:MCFhisto} below
 compares the accretion rate histograms of the full sample, and of a
 revised sample which includes only objects at $z\leq 0.29$.}
\end{figure}
\begin{figure}
\plotone{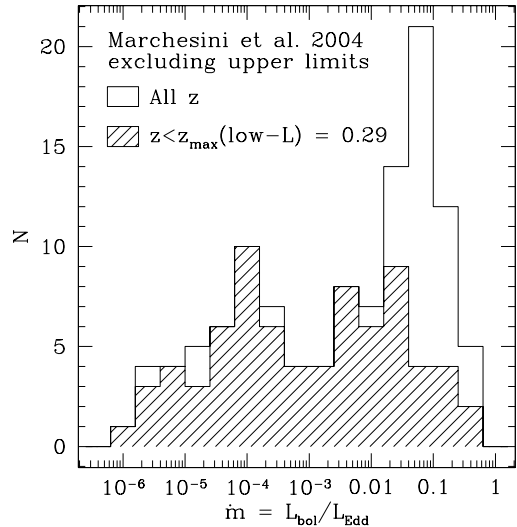}
\caption{\label{f:MCFhisto}Accretion rate histogram for \citet{MCF04}
  objects, excluding objects with upper limits on the bolometric
  luminosity.  The open histogram shows the histogram of \mdot\ for
  the entire remaining sample, while the shaded histogram shows the
  histogram for objects at $z\leq0.29$, which is the highest redshift
  of any source with $\mdot \la 2\times10^{-3}$.  Thus, the shaded
  histogram excludes those sources located in the volume in which
  low-\mdot\ sources are inaccessible to observations.  This
  ``equal-redshift-range'' sample does not show a pronounced gap in the
  accretion rate distribution any more. }
\end{figure}
The selection effects of the \citet{MCF04} sample of objects are the
simplest, as they use mostly objects from two complete flux-limited
radio surveys, the 3CR and B2 surveys.  The 3CR \citep{LRL83,SMAea85}
is a 178\,MHz survey of 13,920 square degrees of sky to 10.9\,Jy. It
includes a mix of low-power FRI (radio luminosity at 178\,MHz in the
range $10^{24} < \Lose < 10^{28}$\,W\,Hz$^{-1}$) and higher-power FRII
radio galaxies ($10^{25} < \Lose < 10^{28}$\,W\,Hz$^{-1}$) at $z\leq
0.3$, as well as radio-loud quasars at $z\geq 0.3$, with luminosities
in the range $10^{27} < \Lose < 10^{29}$\,W\,Hz$^{-1}$.  The B2 survey
\citep{CFFea75_III} is a survey of about one-half the area of the 3C
with a flux limit of 0.2--0.25\,Jy and a magnitude limit $m_V<16.5$ on
the galaxies identified optically with the radio sources, i.e., it
samples a smaller volume than the 3C.  Therefore, it predominantly
includes low-luminosity objects (high-luminosity objects have a lower
volume density and therefore require a large-volume sample to be
found): the B2 objects used by \citet{MCF04} are all at $z\leq 0.15$
and have $10^{22} < \Lose < 10^{26}$\,W\,Hz$^{-1}$.  The sample is
augmented by further radio-loud quasars in the redshift range $0.15
\leq z \leq 0.3$.  Thus, the total sample has substantial overlap in
\Lose\ and redshift.  However, Figure\,\ref{f:MCFmdotz} shows that the
majority of high-accretion rate objects are drawn from a completely
different volume than all the low-accretion rate objects.
Low-accretion rate objects, in particular the low-luminosity radio
galaxies from the B2 sample, are exclusively found at $z\leq 0.29$,
while \emph{all} of the 3C radio-loud quasars are at $z > 0.3$. The
only high-accretion rate objects at $z < 0.3$ included in the
\citet{MCF04} histogram are the 3C radio galaxies (i.e., objects with
both powerful radio jets as well as broad emission lines and some
non-stellar continuum, but with lower optical luminosities than
radio-loud quasars) as well as the radio-loud quasars from
\citet{MD01}.

Figure\,\ref{f:MCFhisto} shows the effect of removing all sources at
$z>0.3$ from the sample, where objects with $\mdot \la 10^{-2}$ are
inaccessible to the surveys used.  The high-\mdot\ peak in the
accretion rate histogram disappears nearly completely, and moves from
$\mdot_\mathrm{peak} \approx 0.1$ to $\mdot_\mathrm{peak} \approx
0.01$.  There is no clear bimodality any more.  Thus, not only is the
gap in the accretion rate distribution not obvious when considering
the full 2-dimensional distribution of objects from \citet{MCF04} in
the \LMplane\ (Fig.\,\ref{f:3cobs}), but the gap is an artifact of
selection effects: high-\mdot\ objects are predominantly found at
$z\ga 0.3$ in the samples used, while low-\mdot\ objects are
restricted to $z\la 0.3$.  Furthermore, even after applying an
identical redshift cut to the samples used by \citet{MCF04}, the
\emph{volumes} sampled to find the different subsets of objects are
still different because of the different areas of sky and that have
been surveyed by the B2 and the 3C survey, and because of the optical
flux limit imposed on B2 galaxy identifications.

\subsubsection{Selection effects in the SDSS quasar sample}

The same criticism (objects with different accretion rates being drawn
from different volumes) also applies, of course, to the objects shown
in Figure\,\ref{f:obs}.  The objects for which \citet{MD04} determine
black hole masses are drawn from the SDSS DR1 quasar catalog
\citep{SFHea03}. This catalog is restricted in the $i$-band, both by
flux limit and by a limit in absolute magnitude $M_i \leq -22$, which
roughly corresponds to $\Lbol \geq 10^{38}$\,W (using the relation
between \Lbol\ and $M_B$ given by \citealt{MD04} and approximating
quasar spectra as a power law with $f_\nu \propto \nu^{-0.5}$,
resulting in $M_B - M_i = 0.3$).  This luminosity cut means that
objects with $\mdot\leq\mcrit = 0.01$ are only included in this quasar
catalog if they satisfy $\Mbh \ga 10^9\Msun$.  The flux limit of the
quasar survey restricts the detection of such objects to $z \la 0.33$
for $ugri$-selected objects ($m_i \leq 19.1$) and $z \la 0.53$ for
$griz$-selected objects ($m_i \leq 20.2$).  The surveyed volume is
therefore limited, and hence the number of such objects that can be
found by the SDSS.

At higher redshifts, the surveyed volume is much larger, but now the
flux limit introduces a redshift-dependent luminosity limit.  As
discussed by \citet{MD04}, the use of the luminosity in the black hole
estimation in a flux-limited implies a correlation between black hole
mass and redshift, leading to an effective lower limit on the black
hole mass as a function of redshift.  Thus, the SDSS quasar survey
misses high-mass black holes with low luminosities.  Furthermore,
\citet{MD04} include only objects with $v_\mathrm{FWHM} \geq
2000$\,km\,s$^{-1}$ in their sample, which excludes objects to the
left and above the dashed line.  This cut is equivalent to a
mass-dependent upper limit on \mdot\ for objects in the sample.

In general, the volume density of faint AGN (more than 3 magnitudes
fainter than those observed by the SDSS) increases towards redshifts
around 2 in the same way as that of luminous quasars \citep{WWBea03},
i.e., faint AGN outnumber bright ones at all redshifts.  We are
therefore missing black hole mass determinations for a substantial
fraction of the AGN population, with a strong bias against
low-luminosity AGN at high redshifts.

\subsubsection{Selection effects for remaining sources}

The remainder of the objects have unquantifiable selection effects,
since they are not drawn from samples with well-defined selection
criteria.  However, \citet{WU02a} give a detailed discussion of
possible selection effects in the construction of a diagram such as
Fig.\,\ref{f:obs}.  The important incompleteness for the present
discussion is again the lack of objects with high black hole masses
($\Mbh \geq 10^8\Msun$) and low luminosities ($\Lbol \leq
10^{38}$\,W).  Since black hole mass correlates with the bulge
luminosity of the host galaxy \citep[][and references therein]{MH03},
a fraction of these objects is missing from AGN samples at all
redshifts: a low-luminosity AGN is more difficult to detect against
the more luminous host galaxy, and the small volume of the
low-redshift universe means that high-mass black holes are rare in it.
There is an additional observer bias towards high-luminosity AGN:
black hole masses are typically first determined for objects drawn
from the brightest samples, such as the 3C radio survey, the Bright
Quasar Survey, or the SDSS quasar survey, even though faint AGN are
more numerous.

Furthermore, there is some natural practical bias towards a certain
type of AGN for a certain method of black hole mass determination. For
example, the use of the stellar velocity dispersion method is only
possible for AGN which are sufficiently nearby and of low luminosity
to allow observation of the host galaxy.  A sample that exclusively
uses this method will therefore be biased against high-luminosity AGN
with low-mass black holes, because these reside in low-luminosity
hosts which are likely to be outshone by the nucleus.  Selection
effects of this kind can be avoided by making use of the broadest
possible range of black hole mass determination methods.

In conclusion, the sample selection function severely influences the
observed distribution of objects in all regions of the \LMplane. In
particular, flux limits on AGN surveys (both explicit flux limits in
complete surveys, and implicit flux limits because the faintest AGN
are practically unobservable) bias any survey against the inclusion of
low-luminosity objects, whose black hole masses are crucial in
detecting the expected difference between efficiently and
inefficiently accreting objects.  Finally, there is a population of
obscured AGN that might be at least as numerous as known AGN
populations \citep{RE04,TUCea04}, but not many black hole mass
measurements exist for Type 2 AGN, and none exist for obscured sources
at high redshift.

Thus, considering distributions like those in Figure\,\ref{f:obs} may
yield a misleading picture because every data point obtains an equal
weight, regardless of the volume that had to be surveyed to find it,
and regardless of the properties of objects that are excluded by
selection effects.  The effect of removing sources at high redshift,
where low-luminosity AGN are not included in flux-limited samples,
from Fig.\,\ref{f:obs} would be similar to the effect on the accretion
rate histogram in that it removes the bulk of objects which are
clearly at $\mdot>0.01$.  The more sound way to perform the test
suggested here will be to determine the volume density of objects with
a given \Lbol\ and \Mbh, preferably in a volume-limited sample.

However, the evidence cited above in favor of the existence of a
transition relied mostly on the asymmetry in the $\mdot$ distribution
of objects with masses from host galaxy stellar velocity
dispersions. Since these are more likely to include low-luminosity AGN
than flux-limited quasar surveys, this conclusion is not affected
severely by the selection effects discussed in this subsection.

\section{Discussion}
\label{s:disc}

The main aim of this paper is to point out that a transition in the
radiative efficiency at some critical accretion rate \mcrit\ should
lead to a change in the distribution of AGN in the \LMplane, and that
this prediction is testable with presently available techniques.  If
the analogy between AGN and black-hole binaries is complete, the
difference between the different accretion states should not only be
evident in the \LMplane, but also manifest itself in a spectral
difference between efficiently and inefficiently accreting
objects. This section considers the spectral evidence for an analogy
of AGN and black-hole binary accretion states, first for the 12
low-\mdot\ objects identifiable in Fig.\,\ref{f:obs} and then for the
remaining objects which are presumably in a high-efficiency accretion
state. I also discuss whether the spectral and jet properties of AGN
in general fit into the same classification scheme as black-hole
binaries.  Before considering these points, I briefly discuss the
evidence for the presence of a transition in the accretion properties
of radio-loud AGN presented by \citet{MCF04}.

\subsection{A bimodality in the \mdot\ distribution of AGN with
powerful jets?}
\label{s:disc.3ctransition}

In \S\ref{s:obs.sel.mcf} above, I showed that the obvious bimodality
in the accretion rate histogram of radio-loud AGN presented by
\citet{MCF04} is to a large degree a consequence of sample selection
effects, with high-\mdot\ objects (high-efficiency accretors) being
drawn from a much larger volume than those with low
\mdot\ (low-efficiency accretors).  This does not, however, imply that
the conclusions drawn by \citet{MCF04} are necessarily invalid ---
\S\ref{s:obs.sel.mcf} merely shows that selection effects are
responsible at least in part for the apparent bimodality, but not that
there is no such bimodality, even though Figure~\,\ref{f:3cobs} shows
no obvious gap in the distribution of radio-loud objects in the
\LMplane.  To clarify this issue, a sample of AGN is needed which
includes \emph{all} objects with \mdot\ in the range under
consideration.

\subsection{Spectral evidence for inefficient accretion?}\label{s:disc.spec.loweff}

We concluded above that the small number of objects in the sample of
AGN shown in Figure~\ref{f:obs} with $\mdot < \mcrit$ at $\Mbh <
10^{9}\Msun$ may be taken as weak suggestive evidence for the presence
of a radiatively inefficient accretion mode.  The 12 objects which
have $\Mbh < 10^{9}\Msun$ and $\mdot < \mcrit=0.01$ are Mrk~3 and 270,
and NGC 513, 1052, 2110, 2841, 3786, 3998, 4258, 4339, 5929, and 6104;
NGC~4258 has a direct black hole mass measurement from water maser
kinematics, while the black hole masses for the remainder have been
determined from the stellar velocity dispersion of the host galaxy and
the $M-\sigma$ relation.  Is there any spectral evidence that these
are in a low-efficiency accretion state?  If the black-hole binary
scheme applied in complete analogy to AGN, the low-efficiency
accretion state ought to show a steady jet, and they should not be
dominated by thermal emission from the accretion disk, i.e., they
should not have a ``big blue bump''.

These objects are all low-luminosity AGN.  Most are classified as
Seyfert galaxies (including both Seyfert 1, 2, and intermediate types)
or low-ionization nuclear emission-line region \citep[LINER;
  see][e.g.]{Hec80}.  Some of the narrow-line objects show evidence
for hidden broad-line regions in polarized light. Mrk~3 and NGC~1052,
2110, and 4258 show clear radio jets \citep{LiuZhang02}, while the
evidence for a radio jet in NGC 2841 is only tentative so far
\citep{NFWea02}.  Thus, about one third of the objects have clear
detections of radio jets. However, the lack of clear jet detections in
the remainder of the objects does not necessarily imply the absence of
jets, but may be due to the lack of suitable observations.  Indeed,
among the \emph{hard}-state galactic BHXBs, only Cygnus X-1 has an
imaged jet, while the presence of jets in other objects is inferred by
indirect arguments, typically lower limits on the source size inferred
from the radio brightness temperature.

\citet{Ho99} considers the SEDs of seven low-luminosity AGN, showing
that the ``big blue bump'' is weak or absent in these. As those
objects are similar to the low-\mdot\ objects identified here, it is
reasonable to expect that the SEDs are also similar, i.e., that the
``big blue bump'' is absent and the objects have a predominantly
non-thermal spectrum, as expected in the analogy to black-hole
binaries.  However, \citet*{SSD04} argue that the lack of an
observable ``big blue bump'' is better explained by absorption
\citep[this possibility had also been discussed by][]{Ho99}.  In
particular, \citet{SSD04} claim that the observed infrared emission
lines and X-ray luminosities argue in favor of the absorption
hypothesis.  They furthermore cite results of detailed photoionization
modeling which favor a black-body ionizing continuum.  Thus, the
evidence for an intrinsic absence of a ``big blue bump'' in
low-luminosity AGN as a class is inconclusive at present.

The spectral energy distributions (SEDs) of two of the
low-\mdot\ objects identified here have been analyzed in detail:
\citet{LACea96} fitted the SED of NGC~4258 with an ADAF spectrum;
however, \citet{FB99} argue that the SED is equally well explained by
jet emission and an ADAF is not necessary, while \citet{YMFea02} argue
on the basis of more recent infrared data that a jet dominates the
emission, while an ADAF still contributes.  \citet*{AUH04} present a
similar juxtaposition of ADAF and jet models for the radio emission in
six other low-luminosity AGN, which have a spectral shape similar to
ADAF models but too high a radio luminosity.  There is a similar
controversy for NGC~1052, for which \citet{GOOea00} propose an ADAF
model, while more recent Chandra observations presented by
\citet{KKRea04} support a jet model.  Regarding the shape of the
optical continuum in this object, \citet{KM83} claim that the narrow
emission-line ratios are best explained with a simple power-law
ionizing spectrum, while \citet{Peq84} argues that a black-body
spectrum with a higher temperature than is usual in AGN may also
account for the observed line ratios. In other words, the same
emission-line ratios can be accounted for by ionizing spectra both
with and without a thermal component.  A similar result is reported by
\citet*{MVG04} who show that different assumptions about the
homogeneity of the narrow-line region can change conclusions from
photoionization modeling about the shape of the ionizing continuum
drastically.  Again, the evidence for the absence of a ``big blue
bump'' is inconclusive. More detailed multiwavelength observations and
comparison to photoionization models are necessary.

\citet{HP01} compute radio-to-optical ratios for low-luminosity AGN
using only optical emission from the nucleus itself (the larger
apertures used in earlier measurements included a large contribution
from starlight).  The resulting radio-to-optical ratios are extremely
large (up to $10^4$), similar to or even exceeding those of radio
galaxies and quasars with powerful jets. This suggests that radio jets
are common in low-luminosity AGN, as predicted by the analogy with
black-hole binaries.  However, while this analogy seems to predict
\emph{both} an SED without strong thermal emission \emph{and} the
presence of jets, observations such as those by \citet{AUH04} and
\citet{KKRea04} seem to imply that these are mutually exclusive when
attempting to account for the radio or X-ray emission of
low-luminosity AGN.  

Thus, it is at present unclear whether low-luminosity AGN such as the
low-accretion rate objects identified in Fig.~\ref{f:obs} actually are
scaled versions of black-hole binaries in the \emph{steady-jet, hard
  X-ray} state.  The best evidence in favor of this hypothesis seem to
be the common scaling relations between X-ray binaries in the
\emph{steady-jet, hard X-ray} state and AGN accreting at low Eddington
ratios, as reported by \citet*{MHD03} and \citet*{FKM04}.  However,
the same scaling relation seems to extend to objects at larger
Eddington ratios as well, only with larger scatter; see Fig.~7 in
\citet{MHD03}. Even a rebinned version of that figure, Fig.~2 in
\citet{MGF03}, shows \emph{steep power-law} (or \emph{very high
  state}) BHXBs and FR~II radio galaxies falling on the same relation
as the low-efficiency sources.  It is somewhat surprising that the
scaling does not break down altogether for the presumed
high-efficiency and super-Eddington sources sources, given that the
innermost regions of the accretion disks are supposed to be in
entirely different physical states in those objects.

\subsection{Spectral evidence for efficient accretion?}
\label{s:disc.spec.higheff}

As a corollary to the situation for low-efficiency accretors, AGN in
the high-efficiency state should be analogues of black-hole binaries
in the \emph{thermal-dominant} state.  When an object enters the
\emph{thermal-dominant} state, the radio emission associated with the
steady jet is quenched \citep{TGKea72}, indicating that there is a
much weaker jet, or no jet at all, in this state \citep{GFP03}.
\citet{MGF03} show that the radio luminosity of AGN with Eddington
ratios placing them in the \emph{thermal dominant} state drops below
the correlation found for sources in the \emph{hard} state (the
correlation is based on a binned version of the data from
\citealp{MHD03} and similar to that found by \citealp{FKM04}), which
is evidence in favor of such an analogy.  A similar suggestion
has been made by \citet{GC01}, who show that the line separating
low-power FR~I and high-power FR~II sources in a plot of AGN radio
luminosity against host galaxy optical luminosity can be interpreted
as a threshold in accretion rate separating the two populations.
However, \citet{CR04} show that at least one-third of the FR~I sources
in the 3C catalog must be accreting in the radiatively efficient
regime to account for their radio luminosities, if their jets are to
be powered by the Blandford-Znajek mechanism.  As mentioned above,
they ascribe the low optical core flux of these sources to absorption.

We can assess whether the presence of jets is restricted to objects in
the inefficient regime by considering Figure~\ref{f:3cobs} again. What
matters to the analogy between black-hole binaries and AGN is that
\emph{all} objects plotted in Figure~\ref{f:3cobs} show extended radio
emission powered by jets.  In the BHXB-AGN unification picture, none
of the objects with $\mdot>\mcrit$ ought to show steady jets. Of the
radio-loud quasars and radio galaxies from the 3C survey, at least the
radio-loud quasars are obviously not in the inefficiently accreting
regime in Figure~\ref{f:3cobs}.  These objects are the most luminous
radio sources in the universe. As jet kinetic power correlates nearly
linearly with low-frequency radio power \citep{Wileta99}, these
sources also have the most powerful jets in the universe. This is
inconsistent with the expectation that the jets should be quenched
\citep[even though the quenching appears to be less extreme in AGN
  than in BHXB, the AGN quenching may have been underestimated due to
  systematic errors in luminosity and black hole
  measurements;][]{MGF03}.  I now discuss a possible resolution of
this contradiction.

\subsection{Are some radio-loud AGN in a ``steep power-law'' state?}
\label{s:disc.thirdstate}

Disks in efficiently accreting AGN do appear to be able to launch
relatively more powerful jets than BHXBs with identical $\mdot$,
apparently implying that BHXB-AGN unification is not perfect.
However, the suggestion has been made that the radio galaxies and
quasars with powerful jets are analogues of black-hole binaries in the
\emph{steep power-law} state with transient X-ray and radio flares,
interpreted as discrete ejections of high-velocity jets
\citep{GFP03,Mei01}. The observations that have most specifically been
taken as evidence for this hypothesis are dips in the X-ray emission
from the blazar 3C\,120 linked to the appearance of new radio knots in
its jet \citep{MJGea04}.  A naive scaling of the duration of these
outburst in stellar-mass systems (few days) by a factor of the black
hole mass ratio for AGN ($10^8$--$10^{10}$) results in expected jet
lifetimes of order $10^6$\,y -- $10^8$\,y for these AGN. It is
encouraging that this brackets the typical jet age of powerful radio
sources of $\approx 10^7$\,y \citep{Wileta99}.  However, all the
radio-loud quasars in Figure~\ref{f:3cobs} have $\mdot \approx 0.1$,
i.e., in the standard thin-disk jetless regime.  Alternatively, if the
accretion rates of the radio-loud quasars are systematically
underestimated by a factor of 2--3, these sources may be in the regime
of the \emph{steep power-law} state (indeed, the black hole masses of
some of the high-redshift 3CR radio-loud quasars may have been
underestimated by a comparable factor, see the note at the end of
\S\ref{s:obs.err}).  But even if the accretion rates derived for the
radio-loud quasars are not correct, there are other sources with
powerful jets, but implied accretion rates firmly in the standard-disk
regime.  This suggests that additional factors besides the accretion
rate govern the disk structure and the appearance of jets (as proposed
by \citealp{Mei01}; also compare \citealp*{CF00,LPK03}).  This is
further supported by the detection of extended radio emission with
FR~I morphology around the quasar E1821+643, which is optically
luminous but has very low radio luminosity \citep{BR01}.  Conventional
wisdom has it that optically luminous quasars either have FR~II jets
(if they are ``radio loud'' by either criterion for radio loudness),
or a jet that is so weak that it cannot leave the host galaxy (or no
jet at all).  While this particular object may have FR~I morphology
because of strong jet precession \citep{BR01}, it is possible that
there are many more optically luminous quasars whose extended emission
has not been detected because of a lack of sufficiently deep radio
observations.

The ``fast ejection'' scenario for jets in powerful sources makes
testable predictions about the X-ray spectra of FR~I and FR~II radio
sources. Radio-loud AGN typically have a harder X-ray spectrum than
AGN with similar optical, but lower radio luminosity, i.e., weaker or
absent jets \citep{EWMea94}. This matches the spectrum of BHXBs showing
high-speed ejections in the \emph{very high} state, whose distinctive
feature is in fact a \emph{steep X-ray power-law} \citep[SPL;
see][]{MR04}, i.e., an X-ray spectrum that is softer than that in the
\emph{jet/hard} state (but still harder than in the \emph{thermal dominant}
state).  As these ejections are non-steady, there are a large number
of BHXBs in the SPL state but without strong radio emission.  The
unification scheme thus predicts the presence of AGN with hard X-ray
spectra similar to those of radio-loud quasars, but without powerful
jets.

There is a possible exception to the radio-loud AGN unification scheme
with implications for BHXB-AGN unification. \citet{EH03} argue that
broad-line radio galaxies (BLRGs) with double-peaked emission lines
accrete in the low-efficiency regime.  Their optical luminosity would
then be lower compared to quasars with similar radio luminosity not
because of obscuration, but because of a flow with lower radiative
efficiency.  In this case, the radio galaxies with double-peaked lines
are relatives of \emph{jet/hard}-state BHXBs with steady, low-speed
jets, not of their fellow FR~II radio sources with optical quasar
spectra and presumed high-speed non-steady ejections (jets).  Indeed,
BLRGs with double-peaked lines have narrow-line ratios indicating
lower ionization states than is typical for radio-loud objects.  

This separation into high- and low-ionization jet sources matches the
findings by \citet{WRBea01} who show that is is necessary to invoke a
dual-population scheme for radio galaxies in order to fit the
evolution of the radio luminosity function.  The high-luminosity
population includes all FR~II sources with strong emission lines,
while the low-luminosity population includes all FR~I and those FR~II
sources with weak emission lines, and presumably those BLRG with
double-peaked emission lines.  Based on such dual-population schemes,
\citet{Mei01} had already proposed that there are two subclasses of
jetted AGN with different modes of accretion and jet generation.
Thus, both the double-peaked BLRGs and the dual-population scheme
represent independent evidence for a class of FR~II radio galaxies
with different nuclear properties that can be explained by
lower-efficiency accretion.  It is important, though, that the
luminosity separating the two radio populations with different
evolution is about a factor of 10 higher than the luminosity
separating sources with FR~I and FR~II morphology, i.e., this
separation is \emph{not} the one considered by \citet{GC01}.

If the BHXB-AGN analogy holds up, there should be detectable
differences between the jets in the high- and low-luminosity radio
populations in terms of speeds and lifetimes.  In this context, the
study of radio galaxies with two sets of radio lobes (double-double
radio galaxies) may become important.  These objects are interpreted
as AGN whose activity has been interrupted for a few Myr
\citep{SBRea00} and their study may provide insights into jet
triggering mechanisms.

Another prediction of the BHXB-AGN unification including the third
accretion sate is that there should be AGN equivalents of BHXBs in the
SPL state that do not currently produce ejections, i.e., AGN which are
identical to FR~II radio galaxies and quasars except for emission that
can be ascribed to jets, but which are different from jetless AGN in
the \emph{thermal-dominant} state. These issues deserve further
investigation that is, however, beyond the scope of the present work.

\section{Summary and conclusion}
\label{s:conc}

Evidence is accumulating now that the different accretion states
observed in black-hole binaries \citep{MR04} will analogously occur in
accretion systems in active galactic nuclei, i.e, for a unification of
black-hole binaries (BHXB) and active galactic nuclei (AGN)
\citep{Mei01,MGF03}.  In this paper, I present a simple observational
test for the existence in AGN of the transition between a radiatively
inefficient state and an efficient one thought to occur at a critical
Eddington-scaled accretion rate $\mcrit \approx 0.01$ in BHXBs
\citep{Mac03}.  The test is based on the determining the location of
objects in the \LMplane, and hence their Eddington-scaled accretion
rate.  The underlying distribution of black hole masses and physical
accretion rates is expected to vary smoothly. Therefore, a change in
the accretion efficiency implies that the distribution of objects in
the \LMplane\ should also change at the critical accretion rate.  I
consider whether such a change is detectable in a sample of AGN with
known bolometric luminosities and black hole masses obtained from
\citet{WU02a} and \citet{MD04}.  I obtain the following results:
\begin{enumerate}
\item The bulk of objects considered here lie in the radiatively
  efficient regime $\mdot > \mcrit=0.01$ of the
  \LMplane\ (Figure~\ref{f:obs}).  Even though the lack of objects at
  $\mdot < \mcrit$ with $\Mbh > 10^9\,\Msun$ is mainly due to
  selection effects, the small number of objects with $\Mbh <
  10^9\,\Msun$ and $\mdot < \mcrit$ is weak suggestive evidence that
  the density of objects decreases below the line $\mdot=\mcrit$.
\item Selection effects are important in shaping the observed
  distribution of objects in the \LMplane, in particular those
  selection effects arising in flux-limited samples
  (\S\ref{s:obs.selection}).  The most severe bias is the lack of
  low-luminosity objects at high redshifts from existing black hole
  mass surveys.  These objects outnumber luminous AGN at all redshifts
  and likely have a different distribution of accretion rates than
  low-luminosity objects in the local universe.  In particular, the
  apparent bimodality of the accretion rate distribution of AGN with
  powerful jets \citep{MCF04} is predominantly due to this selection
  against low-luminosity high-redshift AGN.  The bimodality is
  strongly reduced if high-accretion rate objects are restricted to
  lie at $z<0.29$, the maximum redshift of the low-accretion rate
  objects in the \citet{MCF04} sample.
\item Without a reliable method to determine an individual black
  hole's spin and hence the fraction $\eta$ of binding energy
  liberated by accretion, all determinations of accretion rates may be
  uncertain by a factor of three or greater (from $\eta=0.42$ for
  maximally co-rotating black holes to $\eta=0.06$ for non-rotating
  black holes, or even less for counter-rotating black holes; see
  \S\ref{s:obs.assumptions}).  This and similar uncertainties in the
  determination of black hole masses and bolometric luminosities
  mainly blur the expected change in object density at $\mdot=\mcrit$
  (\S\S\ref{s:obs.err}, \ref{s:obs.lum}).  The systematic effects
  associated with the use of a single-band flux and a bolometric
  correction factor are potentially more seriously distorting the
  apparent distribution of objects, calling for multiwavelength SEDs
  to be determined for a larger sample of AGN.
\item The objects identified as low-efficiency accretors in
  Figure~\ref{f:obs} are all low-luminosity AGN
  (\S\ref{s:disc.spec.loweff}).  The evidence for jets and against the
  presence of thermal ``big blue bump'' emission from these objects is
  inconclusive on close scrutiny, although low-luminosity AGN have
  radio-to-optical ratios similar to those of powerful radio galaxies
  and radio-loud quasars with jets if only their nuclear fluxes are
  considered \citep{HP01}.  Jet and ADAF models are often pitted
  against each other in the literature to explain the radio and X-ray
  emission from low-luminosity AGN, while the unification would
  predict a simultaneous contribution from both \citep[as obtained by
    some authors, see][e.g.]{UH01,YMFea02}. Furthermore,
  photoionization models can account for the narrow-line emission from
  these objects by an ionizing continuum both with and without a
  thermal component \citep{MVG04}. Hence, the apparent lack of a big
  blue bump does not necessarily imply the absence of thermal emission
  from the accretion disk, but could also be due to absorption.  On
  the other hand, broad-line radio galaxies with double-peaked
  emission lines may be a class of objects that is genuinely in the
  low-efficiency regime \citep{EH03}.
\item The energy output of some or all low-efficiency BHXBs is
  dominated by the kinetic energy of jets, so that observational tests
  for an equivalent kinetic energy output from low-luminosity AGN
  should be a powerful test of BHXB-AGN unification, e.g., by looking
  for evidence for deposition of the jet's kinetic energy in the AGN's
  surroundings.  
\item Many objects in the 3C sample have both the most powerful radio
  known radio jets and accretion rates placing them firmly in the
  radiatively efficient regime (Figure~\ref{f:3cobs} and
  \S\ref{s:disc.spec.higheff}), where jet production should have been
  quenched according to the BHXB-AGN unification scheme \citep[as
    noted by][]{Mei01}.  These objects might be the counterparts of
  BHXBs in the \emph{steep power-law} (SPL) state with non-steady fast
  jets \citep[as suggested for 3C\,120;][]{MJGea04}.  Double-double
  radio galaxies, in which the jets have been interrupted for a few
  Myr \citep{SBRea00}, perhaps constitute further evidence for such an
  analogy.  In these objects, jet production is obviously triggered by
  factors other than the accretion rate. Similarly, there should be
  AGN counterparts of SPL BHXBs which are not currently undergoing
  ejections.
\end{enumerate}

A more accurate determination of the distribution of objects in the
\LMplane\ clearly depends on more accurate black hole mass
measurements. The ideal sample to avoid selection effects would be a
volume-limited AGN catalog with complete black hole mass
identification.  This would allow a determination of the accretion
rate distribution in a manner similar to the computation of luminosity
functions (determining volume densities instead of counting
incidences).  Modern surveys like the Sloan Digital Sky Survey's
spectroscopic quasar survey find quasars with a much broader range in
optical colors, i.e., optical SED shapes, than the Bright Quasar
Survey \citep{SG83}. Therefore, more multiwavelength SEDs are needed
to obtain an accurate measurement of \Lbol\ for SDSS quasars, removing
the substantial uncertainty associated with extrapolation from optical
to bolometric luminosity. Since the X-ray emission is particularly
important to test the spectral equivalence of AGN and BHXB accretion
states, a sample of normal AGN with rest-frame hard X-ray spectra,
deep VLA imaging to search for extended radio emission like that
detected around E1821+643 \citep{BR01}, and well-determined black hole
masses would be particularly valuable to confirm that quasars are the
counterparts of jetless BHXBs in the \emph{thermal dominant} state.
Clearly, the simplified theoretical treatment in \S\ref{s:theory}
should be refined by using more detailed relations between the
radiated flux in different parts of the spectrum and \mdot\ for
different accretion scenarios. In particular, this will allow to
explore possible relations between radio loudness and accretion mode
(see \citealp{MGF03} and \citealp*{WHS03}). In a broader context, the
observed joint distribution of luminosities and black hole masses can
be related to a plausible underlying distribution of accretion rates
and resulting radiative efficiencies for specific models of accretion,
to test their applicability to the AGN population as a whole
\citep{Mer04}.

\acknowledgements

This work was supported by the U.S.\ Department of Energy under
contract No.\ DE-AC02-76CH03000.  I am grateful to Ross McLure for
providing the luminosity and black hole mass measurements from
\citet{MD04} in electronic form, to Arieh K\"onigl for a critical
reading of the manuscript, and to Sebastian Heinz for a useful
discussion.  I am particularly grateful to the anonymous referee for
providing rapid reviews which have led to substantial improvements in
the paper.

Funding for the Sloan Digital Sky Survey (SDSS) has been provided by
the Alfred P. Sloan Foundation, the Participating Institutions, the
National Aeronautics and Space Administration, the National Science
Foundation, the U.S.\ Department of Energy, the Japanese
Monbukagakusho, and the Max Planck Society.\footnote{The SDSS Web site is
\url{http://www.sdss.org/}.}

The SDSS is managed by the Astrophysical Research Consortium (ARC) for
the Participating Institutions. The Participating Institutions are The
University of Chicago, Fermilab, the Institute for Advanced Study, the
Japan Participation Group, The Johns Hopkins University, the Korean
Scientist Group, Los Alamos National Laboratory, the
Max-Planck-Institute for Astronomy (MPIA), the Max-Planck-Institute
for Astrophysics (MPA), New Mexico State University, University of
Pittsburgh, University of Portsmouth, Princeton University, the United
States Naval Observatory, and the University of Washington.

\appendix

\section{Systematic errors in the determination of black hole mass and
accretion luminosity}

\subsection{Errors in black hole mass and bolometric luminosity}
\label{s:obs.err}

The bulk of the objects in Fig.\,\ref{f:obs} have virial black hole
masses obtained from measurements of the width of broad lines.  The
largest systematic error for these comes from the uncertain orbital
geometry of the broad line region emitters.  \citet{MD04} prefer to
assume a disk geometry for the BLR, which increases black hole masses
by a factor of 3 compared to assuming random orbits.  Different
assumptions can increase the resulting \Mbh\ by another factor of up
to 150 \citep{Kro01}. A global error in the orbital geometry would
move the entire set of points with masses from reverberation mapping
or the BLR size-luminosity relation horizontally in Fig.\,\ref{f:obs}.
Thus, if all other observables were known to arbitrary precision, the
assumed orbital geometry would change the value of \mcrit\ inferred
from a plot such as Fig.\,\ref{f:obs}.  If the orbital geometry varies
from object to object (but without systematics with respect to black
hole mass or luminosity), this will introduce scatter into the
determined black hole masses.  Like other sources of scatter discussed
below, this would lead to a smoother transition of the object density
above and below $\mdot=\mcrit$ and thus make the threshold harder to
detect.

In addition to a velocity measurement, the virial mass determination
requires knowledge of the size of the BLR.  \citet{MD04} obtain both
the size of the BLR and the bolometric luminosity from the same
measurement of an object's continuum luminosity at either 3000\,\AA\
or 5100\,\AA, using a bolometric correction of 9.8 and 5.9,
respectively, determined from an average quasar spectral energy
distribution (SED) from \citet{EWMea94}.  This implies a correlation
between the bolometric luminosity and the black hole mass in the sense
of the line in the upper left-hand corner of Fig.\,\ref{f:obs}.
Objects move along this line if there is an error in the determination
of the optical luminosity.  Since the line is nearly parallel to lines
of constant \mdot, random errors in this luminosity determination do
not affect the detectability of a threshold in \mdot\ much.

But the use of a constant bolometric correction factor could be a
source of black hole mass-dependent errors in the inferred luminosity
because the optical emission comes from the ``big blue bump'', which
is ascribed to thermal emission from the accretion disk
\citep{Shi78}. It therefore has a temperature and hence a spectral
shape that varies with black hole mass, introducing a systematic error
in the luminosity.  It is also relevant that the SDSS quasar survey
has very different selection criteria from previous optical surveys,
and may therefore include objects with SEDs different from the average
SED determined from earlier quasar samples.  Thus, there is likely
some scatter of the true bolometric correction factor from object to
object.  Either change in the correction factor influences the
bolometric luminosity, while the black hole mass remains
unaffected. This moves objects vertically in Fig.\,\ref{f:obs}.  The
random scatter would blur the transition region in the same way as the
scatter in the orbital geometry discussed above.  A systematic error
with black hole mass would shear the distribution of objects.  This
would still leave the threshold visible, but would lead to a threshold
line that is rotated or shifted with respect to the line $\mdot=\mcrit
= \mathit{const.}$ expected here.  Thus, it would probably not affect
the conclusion about the presence of a threshold, but its
interpretation.

There is the additional possibility, even prediction, that the SEDs of
AGN in the presumed two accretion states differ, just as the states of
X-ray binaries are differentiated by their SEDs.  It is thus likely
that the bolometric correction factor differs between the radiatively
efficient and inefficient objects, so that the calculation of the
bolometric luminosity from the optical continuum luminosity would
\emph{require prior knowledge} of the accretion state and could not be
used to \emph{determine} the accretion state.  The color selection
algorithm for the SDSS quasars does not make any assumptions about the
SED of quasars, but is designed to select all objects that are
outliers from the stellar locus in color space.  It is therefore in
principle possible that it selects AGN in the radiatively inefficient
state, as long as they pass the flux limit. However, a principal
component analysis of the spectra of SDSS quasars shows that the
average spectrum (the highest-order principal component) does include
the big blue bump \citep{Yipea04,YCVBea04}.  The variation from this
average spectrum is dominated by differences in the host galaxy
contribution and the slope of the power law describing the blue part
of the spectrum.  Thus, the vast majority of SDSS quasars must have a
big blue bump, and the use of a single correction factor is in this
respect adequate for this sample.

The objects from \citet{WU02a} with black hole masses determined by
other methods are not affected by the systematic problems discussed so
far, but have some intrinsic scatter. As the various methods often
cannot be cross-calibrated, it is difficult to obtain an accurate
determination of the intrinsic scatter of these methods. The
dispersion in black hole mass is usually estimated in the range
0.3\,dex for masses from the $\Mbh$-$\sigma$ relation up to 0.5\,dex
for the virial methods. The effect of this scatter is again to blur
the transition between objects in the two accretion states.

\citet{MCF04} determined black hole masses for all objects from the
scaling relation between the bulge luminosity of the host galaxy and
the black hole mass.  They quote an uncertainty of 42\% for these
masses, with the same consequences of blurring the transition as for
the other \Mbh\ uncertainties.  There may also be systematic errors
for the high-redshift objects because the stellar populations of the
high-redshift galaxies hosting the 3C AGN are different from those of
the low-redshift galaxies for which the scaling relations were derived
\citetext{Jong-Hak Woo, \emph{priv. comm.}}.  These would make the
black hole masses of the high-redshift objects overestimates by about
a factor of 0.3--1\,dex. This would also explain the absence of
objects at $\mdot > 0.3$ from Fig.\,\ref{f:3cobs}.

\subsection{Measures for accretion luminosity}\label{s:obs.lum}

The change in the distribution of objects in the \LMplane\ could be
undetectable in Fig.\,\ref{f:obs} if the bolometric luminosity is not
a good measure for the accretion luminosity, so that the actual
accretion state is not in fact revealed by the bolometric luminosity.
One problem is clearly that optical, ultraviolet, and soft X-ray
emission is subject to absorption by dust close to the active nucleus,
and the amount of absorption most likely depends on the viewing angle
of the accretion disk \citep[this orientation-dependent obscuration
forms the basis of unified schemes for AGN; see][]{UP95}.  Therefore,
it might be more appropriate to use a measure of luminosity that does
not suffer from obscuration to determine the \emph{emitted} bolometric
luminosity, instead of extrapolating the \emph{received} flux to the
full solid angle by ignoring the obscuration altogether.

The regions least affected by absorption are hard X-rays and the mid-
and far-infrared. Hard X-ray and mid-infrared fluxes (as observed by
the ISO and IRAS missions) of AGN correlate, albeit with a scatter of
up to one order of magnitude ascribed to non-AGN contributions to the
mid-IR flux \citep{LMSea04}, although the bolometric luminosity is
proportionally less affected.  Similarly, the fraction of luminosity
emitted in the far-infrared varies strongly as function of dust
mass. Therefore the hard X-ray luminosity would seem most appropriate.

Indeed, often the X-ray luminosity alone is used as a measure of
accretion luminosity, because the X-ray emission is believed to arise
in the innermost part of the accretion system. However, the scaling of
the disk temperature with black hole mass implies that the thermal
emission from black-hole binaries peaks in the X-rays, while the
thermal emission from AGN disks peaks in the optical/UV (the ``big
blue bump'').  On the other hand, the observed variability in both
wavelength regions implies that the optical/UV emission in AGN is
reprocessed X-ray emission \citep*{UMU97,KB99}, or that the X-rays are
comptonized optical/UV photons \citep{UMPea00,CRBea00}.  Either way,
X-ray and optical/UV emission are correlated and thus measure the same
accretion luminosity.  However, even bolometric corrections from the
X-rays can vary by a factor of a few \citep{RE04}.  Thus, the AGN flux
in \emph{any} wavelength region is a measure of the accretion
luminosity, but with varying amount of scatter.  Additionally, if the
\emph{shape} of the AGN changes with the accretion state, the
bolometric correction cannot be known in advance, as discussed
above. To minimize the impact of these uncertainties associated with
extrapolating the luminosity in a single waveband, summing the
observed infrared, optical/UV and X-ray luminosity is strongly
preferred, and a true bolometric luminosity is the preferred
approximation to the accretion luminosity.

The radio luminosity always makes a small contribution to an object's
bolometric luminosity and can safely be neglected.  However, in those
AGN with strong jets, the kinetic power of the jet also needs to be
provided by the accretion mechanism and should therefore be included
in the energy budget.  This is particularly important since jets and
radio lobes radiate only a small fraction of the jet's kinetic power.
Therefore, the accretion luminosity should ideally be determined as
the sum of the radiative bolometric luminosity of the disk system and
the kinetic power of the jets.  But \citet{Wileta99} concluded that
the fraction of total accretion energy carried away by the jets of
powerful radio galaxies and quasars ranges from only 5\%--50\%. Hence,
the inclusion of jet kinetic power would not change the total power
output of these objects significantly from the radiative \Lbol\ in
Fig.\,\ref{f:obs}, although the accretion rate would be underestimated
slightly, but systematically.  In any case, most objects in
Fig.~\ref{f:obs} are already in the high-efficiency regime by virtue
of their radiative luminosity, so their classification would not
change by inclusion of their jets' kinetic power.

The situation is more complicated for objects in the low-efficiency
regime. The energy output for some, if not all, X-ray binaries in the
\emph{hard} (low-efficiency) state is dominated by kinetic energy
channeled into the jets (\citealp{FGJ03}; \citealp*{GFP03,MMF04}).
The radiative luminosity will then be only a small fraction of the
accretion luminosity.  However, as long as the overall scaling of the
radiative efficiency with accretion rate is not different from what is
assumed here (Equation~\ref{eq:valeps}) --- in other words, as long as
the jet is powered by accretion energy that would otherwise be
advected into the black hole ---, the accretion rate as inferred from
the radiative luminosity will still be correct.  Assuming a complete
analogy between black-hole binary and AGN accretion systems has
further interesting implications for the presence or absence of jets
in AGN as a function of accretion state, which is discussed in
\S\ref{s:disc.spec.higheff}.

In summary, the radiative bolometric luminosity is the adequate
measure of total accretion luminosity.  To avoid the uncertainties
associated with using a single bolometric correction factor, it would
be desirable to obtain true multiwavelength SEDs for a representative
sample of SDSS quasars.




\end{document}